\documentstyle[epsfig]{mn}

\newcommand{\dd}{\delta}
\newcommand{\g}{\gamma_t}
\newcommand{\gam}{\gamma}
\newcommand{\kk}{\kappa}
\newcommand{\tht}{\theta}

\newcommand{\BB}{{\rm M}_{B}}
\newcommand{\bb}{{\rm m}_{B}}
\newcommand{\be}{\begin{equation}}
\newcommand{\ee}{\end{equation}}

\newcommand{\kunit}{\; h {\rm Mpc}^{-1}}
\newcommand{\runit}{\; h^{-1} {\rm Mpc}}

\title[Galaxy-dark matter correlations applied to galaxy-galaxy lensing]
{Galaxy-dark matter correlations 
applied to galaxy-galaxy lensing: 
predictions from the semi-analytic galaxy
formation models}
\author[]
{Jacek Guzik$^{1}$\thanks{uoguzik@cyf-kr.edu.pl}, 
 Uro\v s Seljak$^{2}$\thanks{useljak@princeton.edu} \\
${}^1$Astronomical Observatory, Jagiellonian University,
	 Orla 171, 30-244 Krak\'ow, Poland \\
${}^2$Department of Physics, Jadwin Hall, Princeton University,
Princeton, NJ 08544}
 
\date{Accepted ---. Received ---}

\begin{document}
\maketitle

\begin{abstract} 

We use semi-analytic models of galaxy formation combined with
high resolution
N-body simulations to make predictions for galaxy-dark matter 
correlations and apply them to galaxy-galaxy lensing.
We analyze cross-correlation spectra between the
dark matter and different galaxy samples 
selected by luminosity, colour or star formation rate. We compare
the predictions to the recent detection by SDSS. We show that
the correlation amplitude and the mean tangential shear depend
strongly on the luminosity of the sample on scales below 
$1 \runit$, reflecting the correlation between the galaxy luminosity and 
the halo mass.
The cross-correlation cannot however be used to infer the 
halo profile directly because different halo masses dominate on 
different scales and because not all galaxies are at the centres of
the corresponding haloes. 
We compute the redshift evolution of the
cross-correlation amplitude and compare it to those of galaxies and 
dark matter.
We also compute the galaxy-dark matter 
correlation coefficient 
and show it
is close to unity on scales above
$r > 1 \runit$ for all considered galaxy types. This would allow one
to extract the bias and the dark
matter power spectrum  on large scales from the galaxy and galaxy-dark matter 
correlations. 
   
\end{abstract}


\section{Introduction}

One of the main goals of modern cosmology is to understand the processes of
structure and galaxy formation in the universe.
It is widely believed that large 
scale structure emerged from small
matter density fluctuations through the gravitational instability process. 
Once the fluctuations became large the process of gravitational 
collapse led to the formation of first dark matter haloes, which 
subsequently merged to form larger and larger haloes. 
Gas initially followed the dark matter until it reached sufficiently 
high densities that it was able to cool efficiently and condense at the centres
of the haloes. Subsequent star formation from the cold gas lead to the creation 
of galaxies. While the broad picture of this scenario is widely accepted 
its details are still poorly understood. 

Observational constraints on the processes of structure and galaxy 
formation can be divided into three categories. In the first are 
observations that depend only on the distribution of the dark matter. 
Examples of these
are velocity flows (e.g. Strauss \& Willick 1995), weak lensing 
(e.g. Mellier 1999) or X-ray temperature function (e.g. Blanchard et al.
2000). 
Another example at high redshift are cosmic 
microwave background anisotropies (e.g. Hu et al. 2000). These direct probes of 
dark 
matter are our best hope to determine the cosmological parameters 
and the clustering of dark matter. In the second category are 
observations that only trace galaxies. Examples are galaxy 
luminosity function, colours, morphologies, clustering etc. These 
observations also depend on the underlying dark matter distribution 
to some extent, 
but the relation is often a complicated one and difficult to 
interpret in the absence of dynamical information. In the 
third category are observations that relate the properties of the
galaxies to those of the dark matter. 
An example is the Tully-Fisher relation (Tully \& Fisher 1977), which relates the
luminosity of the spiral galaxies to their maximal rotational velocity,
which is related to the mass of the halo in which the galaxies 
reside. Observations in this category are specially important 
in relating the process of galaxy formation to the process of
structure formation.
Galaxy-dark matter correlations, which can be measured through 
the weak lensing effects such as galaxy-galaxy lensing, galaxy
foreground-background correlations or galaxy-quasar correlations, 
offer another example that 
fall into this category. The goal of this paper is to analyze these
in the context of the galaxy formation models and make predictions
for such observations.

Physics of galaxy formation remains a complicated and 
poorly understood subject and 
various approaches have been developed to address it. 
Two most commonly used approaches are hydrodynamic simulations
(Blanton et al. 1999; Pearce et al. 1999)
and N-body simulations coupled to the
semi-analytical models  (SAMs;
Benson et al. 1999; Kauffmann et al. 1999; Somerville et al. 1998).
First approach has the advantage of better modelling 
the physics, but is currently limited by the resolution.
Second approach is less ab-initio and therefore may miss
some important physics, but currently has a larger dynamic range. 
In this paper we focus on the dark matter 
and galaxy distribution which is quantified in terms of the
cross-correlation function and the respective cross-power spectrum.
Since we are particularly interested in the galactic scales
which are only poorly resolved with hydrodynamic simulations 
we will use SAMs coupled to the N-body 
simulations. These are still limited in resolution
and so we compare the results with analytic modelling (Seljak 2000),
which under certain reasonable assumptions 
allows one to extend the resolution limit of 
the current N-body simulations.

Theoretical predictions on how galaxies trace mass, how is 
galaxy luminosity related to the surrounding dark matter and how can we
extract these relations from the observations are essential if we are
to take advantage
of the large amount of data available in the near 
future from surveys such as Sloan Digital Sky Survey (SDSS) or 2dF.
SDSS has already obtained the first detection of the galaxy-galaxy lensing 
with only a tiny fraction of the total data (Fischer et al. 2000). 
This has been based
on a rough separation between the foreground and the background galaxy 
sample. In the future this will be done with a better precision
by using the photometric redshift information, which will significantly 
increase the signal to noise. Photometric redshifts will also allow one 
to combine the signal as a function of the physical separation rather
than the angular one. Larger data set will allow one to split the 
galaxies as a function of luminosity, colour or morphology. 
As we show in this paper these give
quantitatively different predictions for galaxy-galaxy lensing,
as well as for
galaxy-quasar correlations and foreground-background galaxy correlations. 
This would allow one to extract the 
information on the environment of different types of galaxies.
 
The outline of the paper is as follows. In Section \ref{two} we
present the formalism, simulations and computational methods.
In Section \ref{three} we present results in terms of the cross-power
spectra and respective cross-correlation
functions for different galaxy samples. We also show
their evolution with redshift.
In Section \ref{four} predictions for galaxy-galaxy lensing are
given and compared with some of 
the existing measurements of the effect (Fischer et al. 2000).
In Section \ref{five} we consider the cross-correlation coefficient dependence
on scale and galaxy type and in Section \ref{six} we address the 
prospects of interpreting the
cross-correlations in terms of the dark matter halo profiles.
Conclusions and 
prospects for future work are presented in Section \ref{seven}.

\section{Formalism and Simulations}    
\label{two}

\subsection{Cross-correlation analysis}
We are interested in the distribution of the dark matter around the galaxies 
of a selected type, averaged over all the galaxies in the sample. 
We can quantify this as an excess of the dark matter density above the 
average as 
a function of the radial separation $r$.
This is described by the galaxy-dark matter 
cross-correlation  function, 
\be
\xi_{\rm g,dm}(r)=\left< \dd_{\rm g}(\vec{x}) 
\dd^{*}_{\rm dm}(\vec{x}+\vec{r}) \right>.
\ee  
Here $\dd_{\rm g}$ and  $\dd_{\rm dm}$ are the overdensities of 
galaxies and dark matter, respectively, and we used the 
continuous description of a random field
(see Peebles 1980 and Bertschinger 1992 
for a discussion of the relation between the continuous and discrete realization 
and interpretation of $\xi$ in terms of the excess probability).
We can define similarly the galaxy correlation function 
$\xi_{\rm g,g}$ and dark matter correlation function $\xi_{\rm dm,dm}$ 
by replacing one $\dd_{\rm g}$ or $\dd_{\rm dm}$ in the above equation.

A quantity we extensively use throughout the paper is the power
spectrum of the random process, which is defined as a Fourier transform of
the correlation function. In the case of the cross-power spectrum we have
\be
 P_{\rm g,dm}(k) = \frac{1}{(2\pi)^3} \int \xi_{\rm g,dm}(r) e^{-i\vec{k}\cdot\vec{r}} d^3r,
 \label{psdef}
\ee
where the 
power spectrum depends only on the module of $\vec{k}$ because of isotropy,
\be
P_{\rm g,dm}(k) \delta(\vec{k}-\vec{k'})=
\left<\tilde{\dd}_{\rm g}(\vec{k}) \tilde{\dd}^{*}_{\rm dm}(\vec{k'}) \right>,
\ee
where $\tilde{\dd}(\vec{k})$ is the Fourier transform of $\dd(\vec{x})$.
Often the dark matter power spectrum is written in terms of
the power contributing to the variance of the density field per 
logarithmic interval in Fourier space
\be
\sigma^2(k)_{\rm g,dm}= 4\pi k^3 P_{\rm g,dm}(k).
\ee
\subsection{Cosmological simulations and power spectrum extraction}

In our work we use GIF high resolution N-body simulations 
carried out by the Virgo collaboration (Jenkins et al. 1998).
Our analysis is constrained to the currently popular 
$\Lambda$CDM cosmological model 
with matter density $\Omega_m=0.3$, cosmological constant 
$\Omega_{\Lambda}=0.7$ and the Hubble parameter 
$H_{0}=70\; {\rm km \; s^{-1}Mpc^{-1}}$. The N-body simulations have
$256^3$ particles, each of mass $1.4 \cdot 10^{10} h^{-1}
M_{\odot}$ in a comoving box of size $L=141 h^{-1}\rm{Mpc}$.
Variance of mass fluctuations 
on a scale of $8 h^{-1}\rm{Mpc}$ was $\sigma_8=0.9$,
in agreement with the observed abundance of clusters (Eke, Cole \& Frank 1996). 
N-body simulations are accurate on scales above the gravitational softening 
scale $r_s = 30 \; h^{-1}{\rm kpc}$ and dark matter clumps with more than 10
particles were identified as haloes. For the most of the present 
work we apply the 
absolute magnitude cut to the galaxy selection which eliminates
the very small haloes from the sample. 

For the distribution of galaxies and their physical properties 
we use mock catalogues from the semi-analytic model
of galaxy formation (Kauffmann et al. 1999a,b). These contain information 
on positions of galaxies, their absolute and apparent magnitude in 
several bands, 
star formation rate and mass contained in stars or gas. They also 
include the information on the halo in which they are placed, 
including its mass and virial radius. Empirically motivated dust 
correction has been applied to model the extinction for the $z=0$ sample.
As shown in Kauffmann et al. 1999a,b (see also Benson et al. 1999, 2000,
Somerville et al. 1998 and van den Bosch 1999)  
these models have been successful in reproducing 
a number of the observational
constraints, including galaxy luminosity function, Tully-Fisher 
relation, pairwise velocity dispersion and galaxy correlation function.  

Despite this success of the SAMs it is clear that 
parameterizing the complex process of galaxy formation 
with a few parameters is an oversimplification. It is 
likely that as the new data become available new parameters will need 
to be introduced to account for this complexity. Moreover, at 
present the number of predictions is small and one
would like to have more independent predictions that can be 
verified with the new observations. Galaxy-dark matter correlations 
as measured through the galaxy-galaxy lensing and magnification bias
provide an example of this kind. These have not been explored so 
far with SAMs or hydrodynamic simulations and could 
provide important constraints for such models.

N-body simulations together with the galaxy catalogues allow us to extract 
the galaxy-dark matter cross-spectra.
To increase the dynamic range of the power spectrum we use the technique
proposed by Jenkins et al. 1998.  Instead of taking the Fourier transform of
the particle distribution in a whole simulation box
we divide it
into $m^3$ cubic boxes where $m=2^n, n=0,1,\dots,6$. We interpolate
all the particles onto the small box and perform a $128^3$ Fourier transform.
We verified that increasing the grid does not
affect the resulting power spectra. This method recovers exactly 
all the modes periodic on scale $L/m$. Assuming these modes are 
statistically representative of all the modes it allows one to exploit the 
entire dynamic range of the simulation.
To perform the FFT on a grid in the real space we use the
Nearest-Grid-Point (NGP) mass assignment scheme (Hockney \& Eastwood 1981), 
which involves the least amount of smoothing. Mass assignment however still 
suppresses the power and we divide each 
Fourier mode 
by a window function suitable for the NGP mass 
assignment (Hockney \& Eastwood 1981). 
The averaging for a given amplitude of $k$ is taken over spherical shells 
in the Fourier space.
For each $n$
the longest modes are sparsely sampled and it is better to use 
the larger box (with $n-1$) to extract those. For these reasons we 
only use dynamic range of 2 from one box to the next.
All of the spectra are corrected for the shot noise term which is due to the
discrete nature of galaxies and dark matter particles. This term 
is negligible for the cross-power spectrum and 
dark matter spectrum because of the large number of dark matter particles, 
except for the galaxy samples with very few galaxies in the case 
of cross-spectrum. 
However, for galaxy power spectra shot noise
dominates on small scales with the amplitude inversely proportional to the
number of galaxies in the sample. For rare galaxy samples such as the
very bright galaxies shot noise dominates already on scales of 
order $1 \runit$. This limits the accuracy of the cross-correlation 
coefficient extraction in such samples. 

\subsection{Shear and magnification derivation}

Gravitational lensing of distant objects by 
intervening mass distribution between the source and the observer 
distorts the size and shape of sources as seen in the image plane.
In the regime of small image distortions (weak lensing regime) observationally 
relevant quantities are 
described by the convergence $\kappa$ and the
two-component shear $\vec{\gamma}=(\gamma_1,\gamma_2)$, which
are related to the projected gravitational potential of the lens $\psi(\tht)$ 
via the following relations
$2\kk(\tht)=\partial_1\partial_1\psi(\tht)+\partial_2\partial_2\psi(\tht)$,  
$2\gam_1(\tht)=\partial_1\partial_1\psi(\tht)-\partial_2\partial_2\psi(\tht)$,  
$\gam_2(\tht)=\partial_1\partial_2\psi(\tht)$, where $\tht$ is a two-dimensional
position in the lens plane. 
The convergence $\kk$ 
and may be regarded as a projected mass density along the line of 
sight (ignoring photon displacement effects, see Jain, Seljak and White 1999 
for justification of this assumption),
\be
\kk={ 3 \over 2 } \left(\frac{H_0}{c}\right)^2  
\Omega_{\rm m}\ \int_0^{\chi_s}g(\chi',\chi_s)\ {\delta \over a} \ d \chi'.
\label{Kproj}
\ee
where $g(\chi',\chi) = r(\chi')r(\chi-\chi') /r(\chi) $
is the radial window function, $\chi$ is the radial distance with 
$\chi_s$ is the distance to the source galaxy, $r(\chi)$ is the 
angular comoving distance (equal to $\chi$ in a flat universe 
discussed here), $H_0$ is the Hubble constant, $ \Omega_{\rm m}$
the matter density and $a=(1+z)^{-1}$ the expansion factor.
If most of the cross-correlation signal 
is associated with the galaxy then we can define the 
comoving surface matter density $\Sigma=\int \rho d \chi $, in terms of which 
$\kk=\Sigma/\Sigma_{\rm crit}$. Here 
$\Sigma_{\rm crit}^{-1}(\chi_l,\chi_s)=
4\pi G r(\chi_l)r(\chi_s-\chi_l)/ar(\chi_s)$ 
is critical surface density, where $\chi_l$ is the comoving
radial distances to the lens. 

The quantity we are most interested in is  
tangential shear $\g(\tht)$ which describes
elongation of images perpendicularly to the line connecting the
image and the lens. It may be expressed as the 2-d shear rotated
to
the frame defined by the image and the lens,
$\g(\tht)=\gam_1(\tht)\cos2\beta+\gam_2(\tht)\sin2\beta$, 
where $\beta$ is the relative position angle between the image and the 
lens. In the weak lensing regime 
shear is directly measurable from the image ellipticities.
The relation between the mean convergence
$\overline{\kk}(\tht)$ inside a given circular aperture of radius $\tht$  
and the mean tangential shear along the aperture boundary 
$\langle \g (\tht) \rangle$ is given by 
(Kaiser 1993, Squires \& Kaiser 1996,                                         
Miralda-Escud\'{e} 1996) 
\be
\langle \g (\tht)\rangle= -\frac{1}{2}\frac{d \; \overline{\kk}(\tht)}{d \ln \tht}.
\label{shear}
\ee

Averaged 
projected matter density can be expressed in terms of the galaxy-dark matter 
cross-correlation function. Here we present both the case where the 
redshifts of lens galaxy and background distorted galaxy are known, 
as well as the case when only their distributions are. 
From the definition of the  correlation function
we have that mean number density $n(r)$ of dark matter particles at 
a distance $r$ from 
a chosen galaxy is proportional to the product of the 
mean number density $\bar{n}$ if they 
were randomly distributed in space and respective cross-correlation
function. This relation may be written as
\be
n(r) = \bar{n} [1+\xi_{\rm g,dm}(r)].
\ee
Thus at an angular separation $\tht$ from a galaxy of a
given type the projected matter density is on average given by    
\be
\Sigma(\tht)=\int \bar{\rho} \left(1 + \xi_{\rm g,dm}(r)\right) d\chi,
\label{sig}
\ee
where the integration is taken along the line of sight 
with $\chi$ the 
radial distance and $\bar{\rho}$ the
average matter density.
Keeping $\tht$ constant requires $r^2=\chi_l^2+
\chi^2-2\chi_l\chi \cos \tht$.
We can then use $\kk=\Sigma/\Sigma_{\rm crit}$ to determine 
$\overline{\kk}$
and $\g$. This relation does not include
the change in the 
focusing strength along the line of sight. A more general expression is
\be
\kk=
\int_0^{\chi_s} {\bar{\rho} \over \Sigma_{\rm crit}(\chi,\chi_s)}
\left(1 + \xi_{\rm g,dm}(r)\right) d\chi.
\ee 
Except for the largest angles the dominant contribution is coming 
from the scales closest to the lens galaxy, in which case the approximation 
in equation \ref{sig} is a valid one. This simplifies the analysis, since
for each pair of foreground and background galaxies we only need to 
compute the corresponding $\Sigma_{\rm crit}$. 

When the source and lens redshifts are known (or at least can be 
estimated with the photometric redshift techniques) the optimal way 
to combine the data requires to calculate for each pair of 
lens and source galaxies the corresponding critical 
density $\Sigma_{\rm crit}(\chi_l,\chi_s)$ and to average over the product 
$\Sigma_{\rm crit}\g$. This can then be related directly to 
the integral of the correlation function in equation \ref{sig}.

Often the redshifts of the sources and lenses 
are unknown and the signal has to be averaged
over the redshift distribution of both the source and the lens galaxies.
This is necessary for example when the data sample is
split into the bright and faint sample (Brainerd et al. 1996, 
Griffiths et al. 1996, Fischer et al. 2000). 
The relation in this case is (Moessner \& Jain 1998)
\begin{eqnarray}
\kappa(\theta)&=& 6 \pi^2 \left(\frac{H_0}{c}\right)^2 \Omega_m \int_0^{\chi_0} d\chi 
W_1(\chi) {f(\chi)
\over a(\chi)} \nonumber \\
&\times &\int dk k \ P_{\rm g,dm}\left(k,\chi\right)J_0[kr(\chi)\theta],
\label{ps}
\end{eqnarray}
where $W_1(\chi)$ is the normalized 
radial distribution of foreground galaxies of a given type and
$f(\chi)=\int_\chi^{\chi_0} g(\chi,\chi')W_2(\chi')d\chi'$ is the average
of $g(\chi,\chi')$ over the radial distribution of background 
galaxies $W_2(\chi')$. 

Having  $\kappa(\theta)$ associated with foreground galaxies of a given type 
we  perform an average in a circular aperture 
\begin{eqnarray}
\overline{\kappa}(\theta)&=& 6 \pi^2 \left(\frac{H_0}{c}\right)^2 
\Omega_m \int_0^{\chi_0} d\chi W_1(\chi) {f(\chi)
\over a(\chi)} \nonumber \\
&\times &\int dk k \ P_{\rm g,dm}\left(k,\chi\right) 
\frac{2J_1[kr(\chi)\theta]}{kr(\chi)\theta},
\end{eqnarray}
thus we receive the mean tangential shear as a function of lens-image
separation 
\begin{eqnarray}
\langle \gamma_t(\theta) \rangle &=& 6 \pi^2 \left(\frac{H_0}{c}\right)^2 
\Omega_m \int_0^{\chi_0} d\chi W_1(\chi) {f(\chi)
\over a(\chi)} \nonumber \\
&\times &\int dk k \ P_{\rm g,dm}\left(k,\chi\right) J_2[kr(\chi)\theta].
\label{gammat}
\end{eqnarray}


\section{Galaxy-dark matter correlations} 
\label{three}

\subsection{Correlations at z=0}

In this section 
we investigate  the cross-power
spectra and the cross-correlation functions for several 
galaxy samples. We choose 
galaxies of different absolute luminosity in 
B-band from $\BB=-18$ at the completeness limit of the sample 
to the very rare bright galaxies with $\BB<-21$.
We also select red galaxies with ${\rm M}_B-{\rm M}_V>0.8$ 
and blue
galaxies with ${\rm M}_B-{\rm M}_V<0.8$
in addition to the absolute magnitude cut. 

\begin{figure*}
\epsfig{file=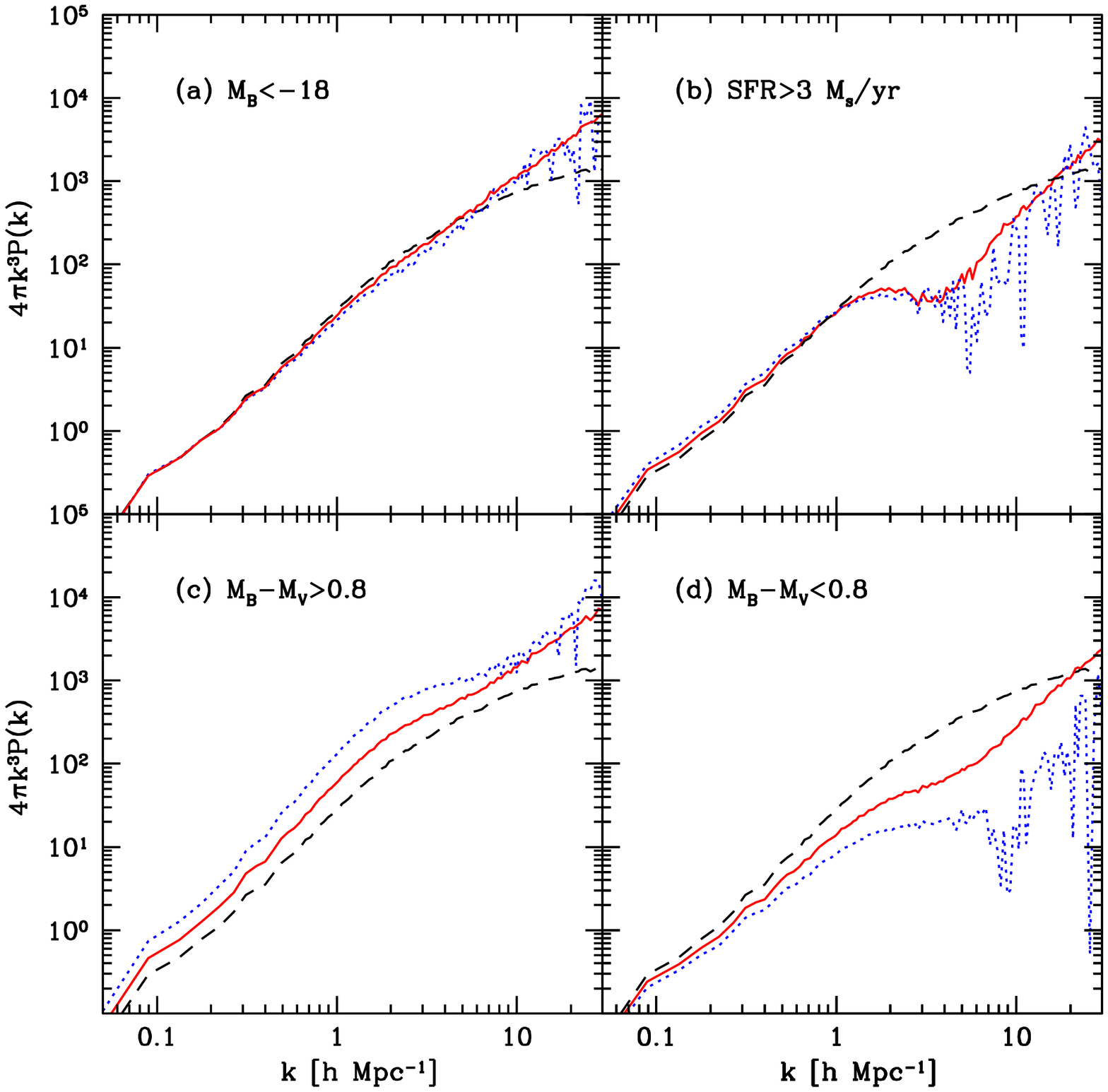, width=150mm}
\caption{
Power spectra of the dark matter (long dashed), galaxies (short
dashed) and galaxy dark-matter cross-power spectra (solid)
for selected samples of galaxies. All power spectra are at
$z=0$.
In panel (a) galaxy selection is $M_B<-18.0$,
in panel (b)  star formation
rate $> 3 {\rm M}_{\odot}/{\rm yr}$, in
panel (c) red colour with $M_B-M_V>0.8$
and in panel (d) blue 
with $M_B-M_V<0.8$, both in addition to $M_B<-18$.}
\label{fig1}
\end{figure*}

Examples of several power spectra for redshift $z=0$ are shown in 
Fig. \ref{fig1}. 
In Fig. \ref{fig1}a we show the power spectra for galaxies 
brighter than $\BB<-18$, for which the sample is complete 
in the sense that most galaxies brighter than this limit form in haloes 
more massive than $1.4\cdot10^{11} h^{-1} {\rm M}_\odot$, which is the 
smallest mass halo that can form in our simulations.
The galaxy power spectrum is a power law 
up to a large $k$, while the dark matter power spectrum is 
continuously changing slope and becoming less steep. 
The galaxies are unbiased on large scales where the two spectra agree.
The 
cross-spectrum has a similar amplitude to the galaxy and dark 
matter spectrum on large scales, as expected if the galaxies are an
unbiased tracer of the dark matter. Cross-spectrum 
grows above both galaxy and dark matter spectrum on smaller scales.
This can be understood with a model 
where for a given halo mass range there are more individual 
galaxies (which contribute to the cross-spectrum) than galaxy pairs
(which contribute to the galaxy spectrum, see Seljak 2000). 
In addition, in a smaller halo
with typically just one galaxy in it the galaxy is at the halo 
centre with dark matter particles distributed as the halo profile. 
In contrast, for the dark matter particles the number of pairs at a 
given separation is suppressed because not all dark matter particles
are at the centre. This suppresses the auto power spectrum of the
dark matter relative to the cross-spectrum.

In Fig. \ref{fig1}a
the transition between the 
large and small scales in the cross-spectrum is continuous
for the sample selected only by absolute magnitude.
Figure \ref{fig1}b shows the same for the star-forming galaxies. 
These are mostly in the field and show a more prominent transition
in the cross-spectrum between the correlations between haloes
for $k<3 \kunit$ and the profile of the haloes themselves at 
$k>3 \kunit$. For galaxies selected only by magnitude groups 
and clusters also contribute to correlations, which fills in the 
transition region between the two (see also fig. 6 in 
Seljak 2000), while for star forming galaxies which are predominantly
in the field groups and clusters do not contribute. 

In Fig. \ref{fig1}c,d red and blue samples of the galaxies 
are shown. Red galaxies are quite biased on large scales, 
indicating that they form predominantly in haloes more massive than the 
nonlinear mass scale (i.e. groups and clusters). On smaller
scales the cross-correlation drops off more rapidly than the 
corresponding spectra in Fig. \ref{fig1}a and b. This is 
because red galaxies in this model typically do not reside at the halo 
centres and thus the power spectrum suffers similar suppression 
as the dark matter. 
In contrast, blue galaxies in Fig. \ref{fig1}d
are unbiased on large scales. The galaxy power spectrum 
is flat on small scales indicating there are very few haloes with 
more than one blue galaxy. Cross-spectrum shows similar features 
on large scales, but begins to rise on small 
scales where the galaxies that are at the halo centres become dominant 
and the cross-spectrum
reflects their own dark matter halo profiles. Overall, the
blue galaxies
show similar features to the star forming galaxies, although the latter 
show more prominent transition between intra and inter halo correlations.

Fig. \ref{fig2} shows the cross-spectra and the 
cross-correlation functions for 
galaxies in narrow magnitude bands in addition to the galaxies in Fig. 
\ref{fig1}. 
On large scales the spectra of narrow-band magnitude galaxies 
agree, indicating that 
the galaxies on average 
reside in unbiased haloes around or somewhat below the nonlinear mass. 
Significant fraction of them reside in groups and clusters, which 
explains the smooth transition between the large and the small scales.
On small scales however there are significant differences in the 
amplitude of correlations, 
which increases with the brightness of the sample. This is 
similar to the Tully-Fisher relation, where the luminosity of 
spiral galaxies is correlated with the maximal circular velocity of 
the disc, which is correlated with the circular velocity of the halo
in the SAMs used here. The 
differences between 
the models are significant already at $100\; h^{-1}{\rm kpc}$, where 
the complications 
because of the baryonic effects on the dark matter distribution 
are negligible. This shows the potential of galaxy-galaxy 
lensing, which can provide independent information 
on the relation between the galaxy luminosity and its dark matter 
environment and can probe regions where baryon effects are 
negligible. Because of the weakness of the signal 
this can only be done statistically for a large sample of 
the galaxies and one must understand 
the complications caused by effects such as the contribution from 
groups and clusters, number of galaxies inside the 
halo as a function of halo mass, central versus satellite galaxies 
inside the halo etc. These are discussed further below.

\subsection{Redshift evolution of cross-correlation}

From the high z catalogs of the galaxy and dark matter positions we extract
the evolution of the galaxy-dark matter correlation function 
for galaxies brighter than $\BB<-19.5$ and
brighter than $\BB<-21$.
No dust correction has been applied to these data so we focus
only on the amplitude of the correlation function at different 
scales as a function of redshift.
Figure \ref{fig7} shows the redshift evolution of the amplitude of the correlation
function at $2 \runit$, $3 \runit$ and
$8 \runit$ for all three spectra. The amplitude of 
the dark matter correlation 
function decreases with redshift because of gravitational instability,
which 
causes density perturbations to grow in time. On the other hand, the 
amplitude of the   
galaxy correlation function first decreases for the same reason, but
then increases again because 
at higher redshifts bright galaxies are only found 
in the rare massive haloes, which are 
biased with respect to the dark matter (Kauffmann  et al. 1999b). 
For brighter galaxies (right panels) the biasing at higher redshift is more 
important and leads to a larger amplitude than for fainter galaxies. 
Amplitude of  the
cross-correlation function falls in-between the amplitudes of 
dark matter and galaxy correlation function. From the 
model with correlation coefficient of unity discussed below 
one predicts it to be the geometric mean, which is 
a good description even for the galaxies that are highly biased, since we are
looking at scales above $1 \runit$, where correlation coefficient 
is close to unity (see next section).
The cross-correlation amplitude typically 
does not increase at higher z, but becomes 
flat. 


\subsection{Galaxy-galaxy lensing predictions}  
\label{four}

In this subsection we present predictions of 
galaxy-galaxy lensing effect for samples of galaxies
discussed in previous section.
For simplicity we model source and lens galaxies as being at fixed redshifts
$z_l=0.16$ and $z_s=0.32$, so that the 
critical surface density corresponds to that of SDSS sample 
(Fischer et al. 2000).
We use $z=0$ dust corrected sample to extract the galaxy-galaxy lensing 
signal, but we verified 
that for a shallow surveys such as the SDSS this assumption does not 
significantly affect the results. 
Results presented in Fig. \ref{fig3} give the 
mean tangential shear as a function of the observed 
angular separation from the lens. 
For the considered cosmological model $10 \; {\rm arcsec}$ 
corresponds to a physical distance in the lens plane of $22.4 \; h^{-1}{\rm kpc}$.
At the redshift $z=0.16$ galaxy of absolute luminosity $\BB=-20$ 
corresponds to 
an apparent magnitude $\bb=19.4$.

\begin{figure*}
\epsfig{file=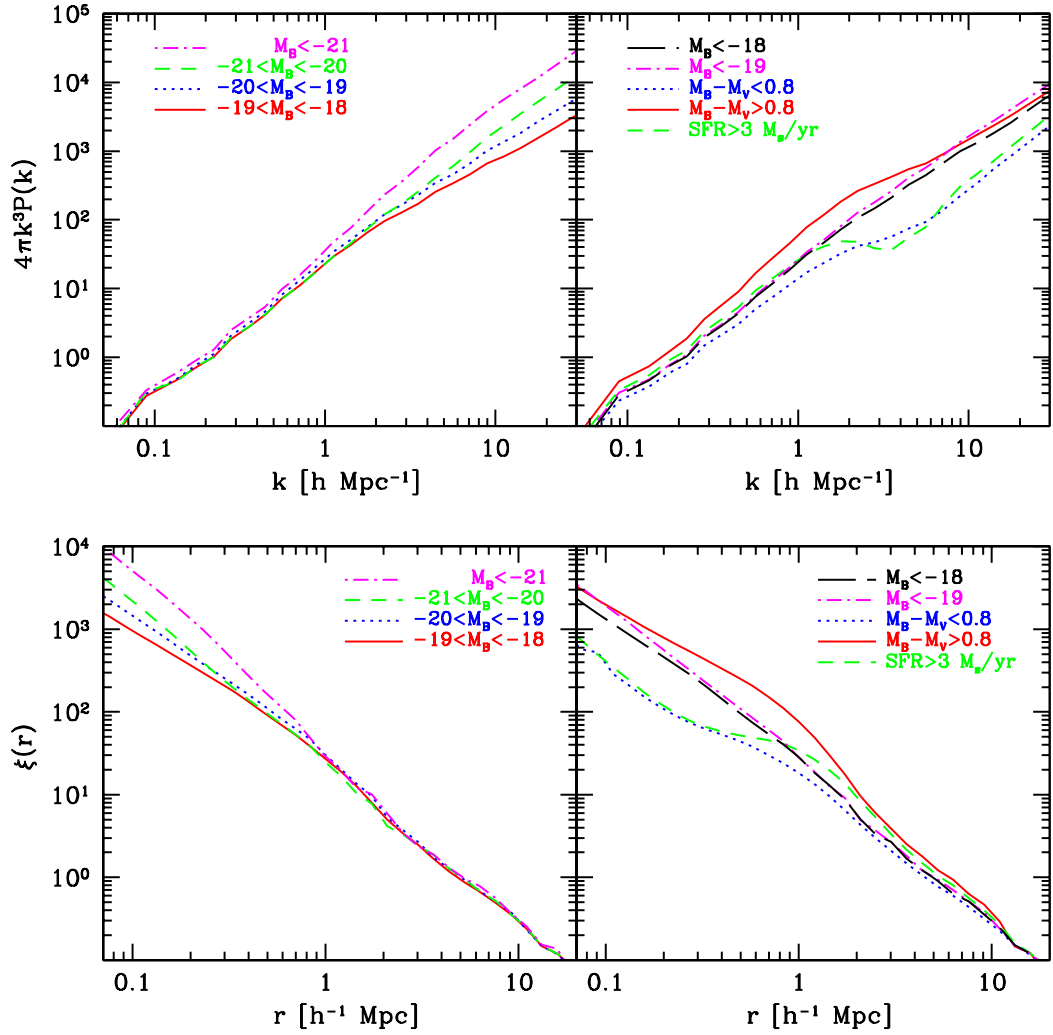, width=150mm}
\caption{
Galaxy-dark matter cross-power spectra and respective cross-correlation
functions for 
luminosity bands (left panel) and colour and star formation rate
(right panel).
In the lower panels the cross-correlation functions are presented for
the same samples as in the upper panel.}
\label{fig2}
\end{figure*}

\begin{figure*}
\epsfig{file=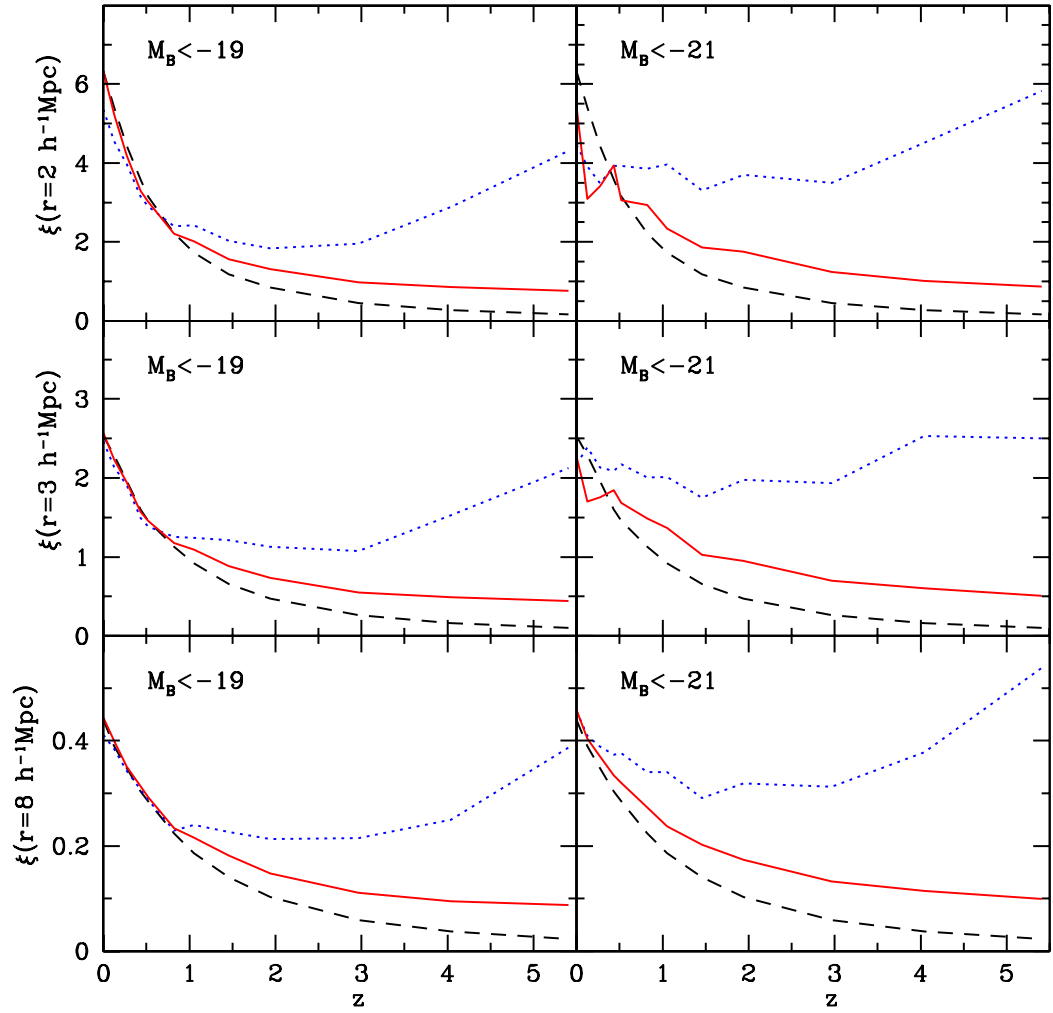, width=150mm}
\caption{
Evolution of the cross-correlation function for two galaxy
samples,
$M_B<-19$ (left panel) and $M_B<-21$ (right panel),
at separation $2 \runit$, $3 \runit$ and $8 \runit$.
Solid line shows galaxy-dark matter cross-correlation amplitude,
dotted galaxy auto-correlation
and
dashed the dark matter auto-correlation.
}
\label{fig7}
\end{figure*}

\begin{figure*}
\epsfig{file=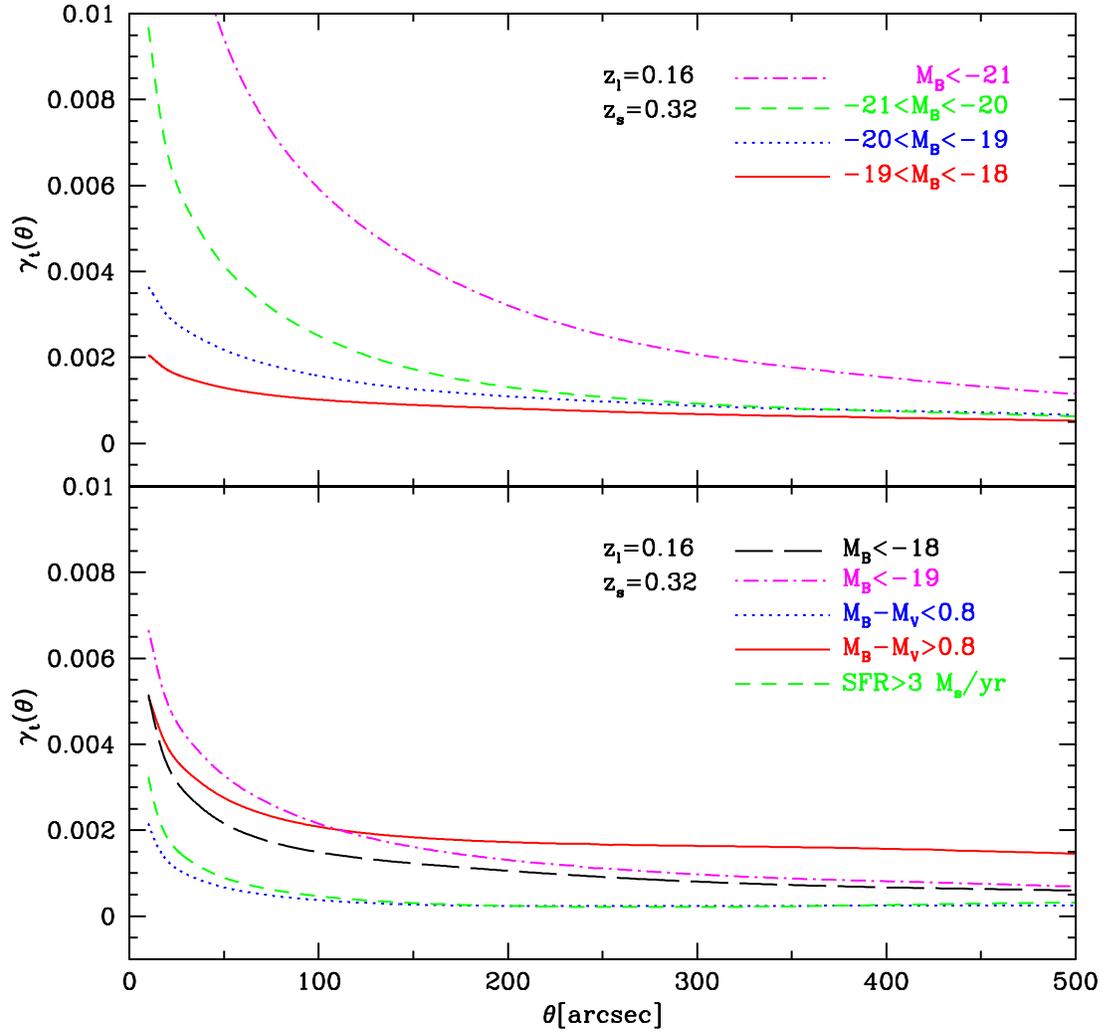, width=150mm}
\caption{
Tangential shear as a function of angular distance from the lens
galaxy for the same models as in figure \ref{fig2}. 
Background galaxies are assumed to be placed at a fixed redshift
$z_{s}=0.32$ and foreground galaxies at a redshift $z_{l}=0.16$ that
mimics the
mean redshift of lenses and sources for SDSS galaxy-galaxy lensing
detection.
}
\label{fig3}
\end{figure*}

Plots in the upper panel of Fig. \ref{fig3} are for the same luminosity
intervals as in Fig. \ref{fig2}. 
For the bright galaxies tangential shear is larger 
than for the faint ones as predicted from the results in Fig. 
\ref{fig2}. Note also that the slope increases with luminosity. 
Fitting the spectra to a power law gives the slopes
varying from $-0.26$ at the faint end to $-0.54$ at the bright 
end, but in general  single power law provides a poor fit to 
the spectra (the slope steepens at larger angles). The slope is considerably 
shallower than the singular isothermal sphere (SIS) distribution with the 
slope $-1$, 
predicting that with better data one should be able to see the deviations 
from SIS. To compare it to the Tully-Fisher relation one can 
compare the predictions for $\gamma_t$ at a given angle, which we 
choose to be $\theta=100 \; {\rm arcsec}$ to maximize the signal to noise in SDSS.
This corresponds to $220 \; h^{-1}{\rm kpc}$ transverse distance from the galaxy. 
Assuming $\gamma_t \propto v_c^2$ where $v_c$ is the circular 
velocity  and fitting this to the relation $L \propto v_c^{\alpha}$ one finds 
$\alpha=3.5$, which is in a good agreement with the Tully-Fisher relation. 
This agreement is however partially fortuitous, because the slope of 
$\gamma_t$ is not well fit with the SIS model and so the  
$L-v_c$ relation must change with the angle at which the comparison is made. 

In the lower part of Fig. \ref{fig3} we show the results for
$\BB<-18$, $\BB<-19$, star forming galaxies, red and blue galaxies. 
Since the luminosity function peaks at $\BB=-19$
galaxies brighter than $\BB<-18$ to $\BB<-19$ 
should be compared to the SDSS 
apparent magnitude selected sample.
It is encouraging that the predictions are 
in a good agreement with the observed signal. 
We should caution that this is not meant to be a quantitative
analysis of SDSS results, because we do not simulate SDSS colour 
bands, selection criteria and galaxy redshift distribution, all 
of which depend also on the luminosity function of the sample 
and not just on the relation between galaxy luminosity and its dark matter
halo environment. With photometric redshift information the dependence on the 
luminosity function can be eliminated and observations will 
provide a more direct constraint on the dynamical environment of 
the galaxies.

\begin{figure*}
\epsfig{file=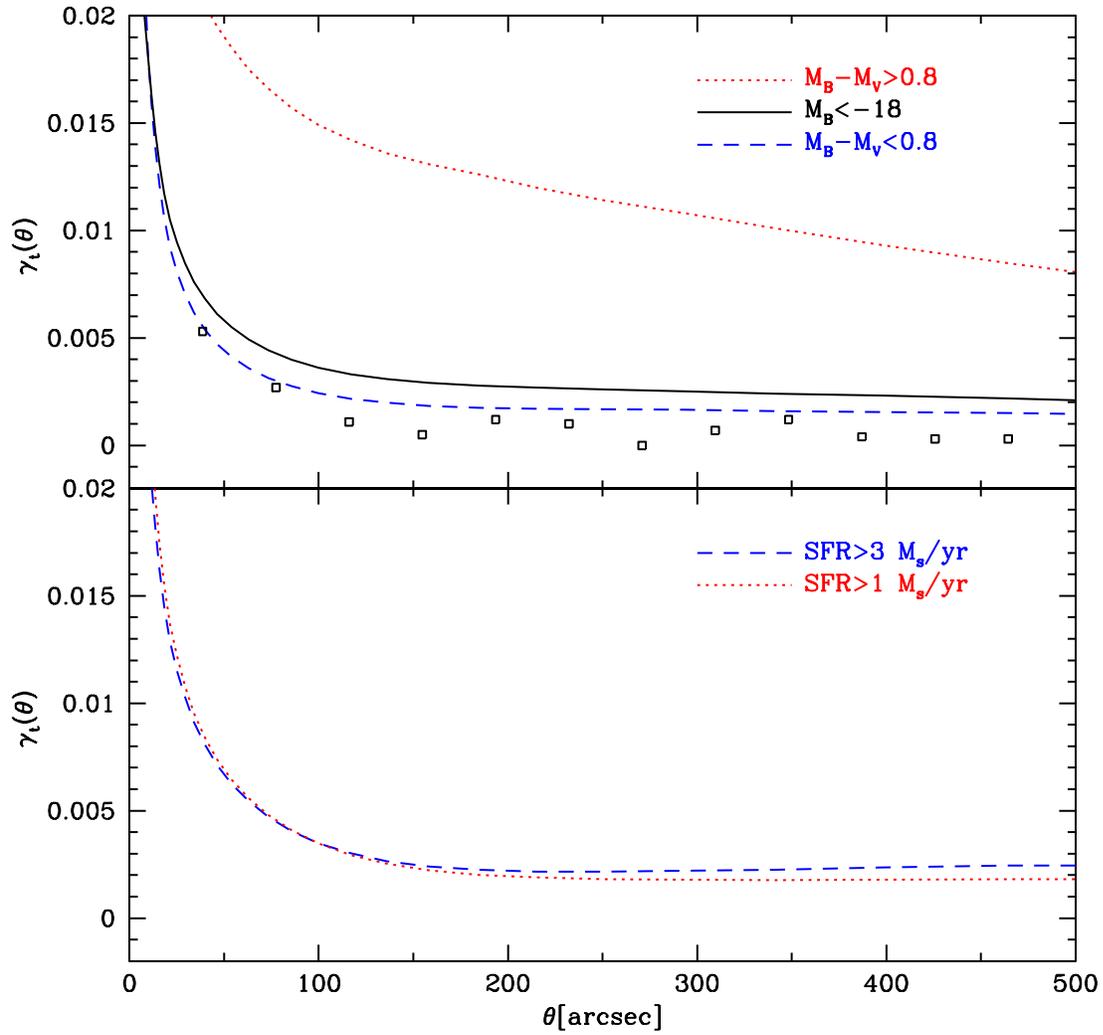, width=150mm}
\caption{
Tangential shear as a function of angular distance. 
Background and foreground galaxies are assumed to be distributed 
as in the SDSS galaxy-galaxy lensing observations. SDSS detection
in the $r'\rm{-band}$ is presented as squares in the upper panel.   
}
\label{fig3a}
\end{figure*}

In Fig. \ref{fig3a} we show mean tangential shear dependence on
the angular scale for red, blue and star forming galaxies, all with the luminosity cut $\BB<-18$.
Also shear for all galaxies brighter than $\BB=-18$ is shown. 
Squares in this figure are observational points from SDSS (Fischer et al. 2000).
Here we include the evolution of 
the cross-correlation function with redshift. Realistic 
lens and source distributions 
provided by the SDSS team
are used in equation (\ref{gammat}). They were derived as approximation of the 
respective SDSS distributions  by the power-law with exponential cut-off of the form
\begin{eqnarray}
W_i(z)= \frac{\beta_i z^2}{(z_i)^3 \Gamma\left(3/\beta_i\right)} 
        e^{-\left(\frac{z}{z_i}\right)^{\beta_i}} 
\end{eqnarray}
with $z_1=0.17$, $\beta_1=2.3$  and $z_2=0.35$, $\beta_2=1.7$ for lens and source samples 
respectively.

Our galaxy samples are roughly complete for galaxies brighter than $\BB=-18$
so we miss some lensing signal from the 
nearby galaxies ($z<0.02$) when we use SDSS
window functions $W_1(\chi)$ and $W_2(\chi)$. But uncertainties related to measured source 
redshift distribution are likely to be more important in this comparison.  

Qualitatively the predictions follow the observations, but we
notice that predicted lensing signal is above that detected by SDSS 
specially on large scales.
One of the reasons for this behaviour is the discrepancy between the 
luminosity function from 
simulations we base on (Kauffmann et al. 1999a) and that obtained from observations
(Lin et al. 1997).
Modelled luminosity function in the $\Lambda CDM$ model lacks less
luminous galaxies, fainter than $\BB=-21$, and has an 
excess of very luminous galaxies, brighter 
than $\BB=-23$, when compared to the present day observations. 
Since as shown in Fig. \ref{fig2} there is a strong luminosity bias 
expected for the cross-correlation amplitude this enhances the
theoretical prediction for the shear amplitude. The other uncertainty 
is already mentioned poorly known redshift distribution function. 
Both of these currently limit the interpretation of results, but
will be eliminated once the photometric redshift techniques become 
realiable. This is specially promising in the case of SDSS where 
5 colour photometry should enable one a very accurate redshift determination. 

\section{Cross-correlation coefficient}
\label{five}

If the bias between the galaxies and the dark matter is not constant, as 
indicated from the results in Fig. \ref{fig1}, then one 
cannot use its determination at a given scale and apply it at another scale.
This complicates the extraction of the dark matter power spectrum from 
the galaxy power spectrum and additional information is 
necessary to break the degeneracy. One can parameterize the 
ignorance with a
scale dependent bias (van
Waerbeke 1998, Dolag \& Bartelmann 1998), where the 
relation between the galaxy, cross and matter 
power spectra is $P_{\rm g,dm}(k)=b(k)P_{\rm dm,dm}(k)$ and
$P_{\rm g,g}(k)=b^2(k)P_{\rm dm,dm}(k)$. In this case from a measurement 
of galaxy and cross-spectrum one can determine both $b(k)$ and 
$P_{\rm dm,dm}(k)$. Even this relation is however not 
general, since the bias factor that relates 
one pair of spectra may not be the same as that for another pair.
One can generalize this by introducing
the cross-correlation coefficient between the 
galaxy and the dark matter power spectrum
\be
r^2(k)= \frac{P_{\rm g,dm}^2(k)}{P_{\rm g,g}(k)P_{\rm dm,dm}(k)}.
\ee 
This definition implies that in the case of a scale dependent but linear
bias discussed above 
the cross-correlation coefficient $|r(k)|=1$. 

In Fig. \ref{fig8} we present 
the correlation coefficient as a function of scale for a set of galaxy
samples selected by luminosity, colour or star formation rate criteria.    
In the right panels we show the correlation coefficient 
without the subtraction of 
the shot noise term arising form the discreteness of galaxies. 
In this case the correlation coefficient is 
constrained to $|r(k)| \leq 1$. More commonly however the shot noise
is subtracted from the galaxy spectrum, which is shown in the 
left panels of Fig. \ref{fig8}.    
In this case for all of the considered galaxy samples the correlation
coefficient remains close to unity up to  
$k \sim 1 \kunit$. 
This is encouraging because
when $r(k)=1$ one can
extract the bias parameter
$b(k)$ and the dark matter power spectrum $P_{\rm dm,dm}(k)$ directly by
measuring the galaxy power spectrum and the cross-power spectrum 
from the galaxy-galaxy lensing. This means that at least on large scales
there is hope that bias could be determined with this method.

The correlation coefficient 
remains closer to unity on small scales  
when the shot noise is subtracted as opposed to 
when it is not. This is related to the 
fact that for a given halo mass 
the pair weighted number of galaxies 
inside the halo
$\langle N(N-1)\rangle^{1/2}$
agrees with the linearly 
weighted number $\langle N\rangle$ 
for large halo masses (Seljak 2000) and that for large haloes 
there are many galaxies inside the halo so the central galaxy, 
if present, does not have a dominant contribution. 
Since the power spectrum on large
(but still nonlinear) scales is dominated by large haloes this gives 
$r(k) \sim 1$. For small haloes typically 
$\langle N\rangle > \langle N(N-1)\rangle^{1/2}$ because 
there are many haloes with just one galaxy in it, which gives
rise to $r(k)>1$ when shot noise is subtracted from the galaxy 
spectrum. In addition, for haloes with just one galaxy this galaxy
is usually at the centre of the halo, which enhances the cross-correlation 
relative to the case where the galaxies are distributed like dark matter 
inside the halo 
(the distinction between the central galaxy and all galaxy 
cross-correlation is discussed further in the next section).
Both of these effects become prominent when $\langle N\rangle <1$ and
so are more important for galaxies that are rarer and/or on smaller 
scales where small haloes dominate.
This agrees with the 
results in Fig. \ref{fig8}, where 
for the brightest and rarest sample of 
galaxies, $\BB<-21$, the correlation coefficient 
increases monotonically already from
$k \sim 0.5 \kunit$, whereas for the faintest and most abundant 
sample $\BB<-18$ 
the cross-correlation coefficient is unity up to $k \sim 3 \kunit$. 

It is interesting to note that for the 
sample of red galaxies (${\rm M}_{B}-{\rm M}_{V}>0.8$ and $\BB<-18$), 
which exhibit a strong scale dependent bias as seen from 
Fig. \ref{fig1}, $r(k)$ is unity over several orders of magnitude in $k$. 
In small haloes  
where $\langle N\rangle \ll 1$ one has
$\langle N\rangle \sim \langle N(N-1)\rangle^{1/2}$ (Seljak 2000).
These galaxies therefore cannot be the 
central galaxies inside the haloes which would give 
$\langle N\rangle > \langle N(N-1)\rangle^{1/2}$. 
For the same reason the cross-spectrum is 
also not enhanced relative to the dark matter, both of which explains 
why $r(k)\sim 1$ over the entire range of scales. While this is 
suggestive we should be cautious 
not to overinterpret these results since SAMs may not be 
accurate in such detailed properties. It will be interesting to examine
whether this prediction can be confirmed from the observations directly.

For galaxies with star formation rate (SFR) above
$3 {\rm M}_{\odot}/{\rm yr}$ the cross-correlation
coefficient $r(k)$ has a strong decrease on scales of order $2-4 \kunit$.
This is 
caused by a decrease in $P_{\rm g,dm}(k)$ (and $P_{\rm g,g}(k)$) 
relative to $P_{\rm dm,dm}(k)$. The reason for this decrease 
is the transition from the correlations between the haloes to correlations 
inside the halo, which is prominent for star forming galaxies which are 
predominantly in the field,
but is washed out by the group and cluster contribution in the 
dark matter power spectrum. 

\section{Dark matter halo profiles} 
\label{six}
In previous sections we have presented the predictions for the
galaxy-dark matter cross-correlations for various galaxy 
selections. We have shown that the predictions strongly 
depend on the galaxy luminosity, colour and star forming rate. 
For example, we observe a strong luminosity dependence with more
luminous galaxies showing a stronger cross-correlation on small
scales because such galaxies reside in more 
massive haloes. These haloes are still below the nonlinear mass
scale and no luminosity bias caused by halo bias 
is observed on large scales, nor is any luminosity bias
observed for galaxy auto-correlation. This shows that 
galaxy-galaxy lensing can be a sensitive probe of halo masses
for different galaxy types. Such interpretation was the 
basis for most of the galaxy-galaxy lensing analysis so far 
(Brainerd et al. 1996, Griffiths et al. 1996 , Hudson et al. 1998). 
Still, the relation between the 
cross-correlation spectrum and the halo mass profile may be a
complicated one and in this section we explore this in more 
detail. 

In the hierarchical clustering picture
galaxies form in the dark matter haloes which subsequently merge into
larger structures. Galaxies which are not members of a cluster or a
group at the time of observation should have more extended haloes than
those of same luminosity already swollen by a cluster because 
tidal forces in a potential well of a cluster tend to strip the galaxies of
their haloes. 
Present observations of the weak gravitational lensing within the
clusters seem to support this picture (Natarajan et al. 1999).   
However, such galaxies are nevertheless 
embedded in larger haloes 
and galaxy-galaxy lensing will be sensitive also to the 
gravitational lensing effect of these more massive and more extended structures.
This effect is specially important on scales above $100 \; h^{-1}{\rm kpc}$
 where the haloes 
of individual galaxies have only a weak signal which can easily be dominated
by a sub-population of galaxies in groups and clusters (Seljak 2000).
This effect is also important on smaller scales, because 
there is a range of halo masses that 
contribute to galaxies selected by their (absolute or apparent) 
luminosity and haloes of different mass 
can dominate on different scales. This makes the interpretation of 
the cross-correlation in term of the halo profile more complicated 
than simply an averaged dark matter profile of a typical galaxy. 

\begin{figure*}
\epsfig{file=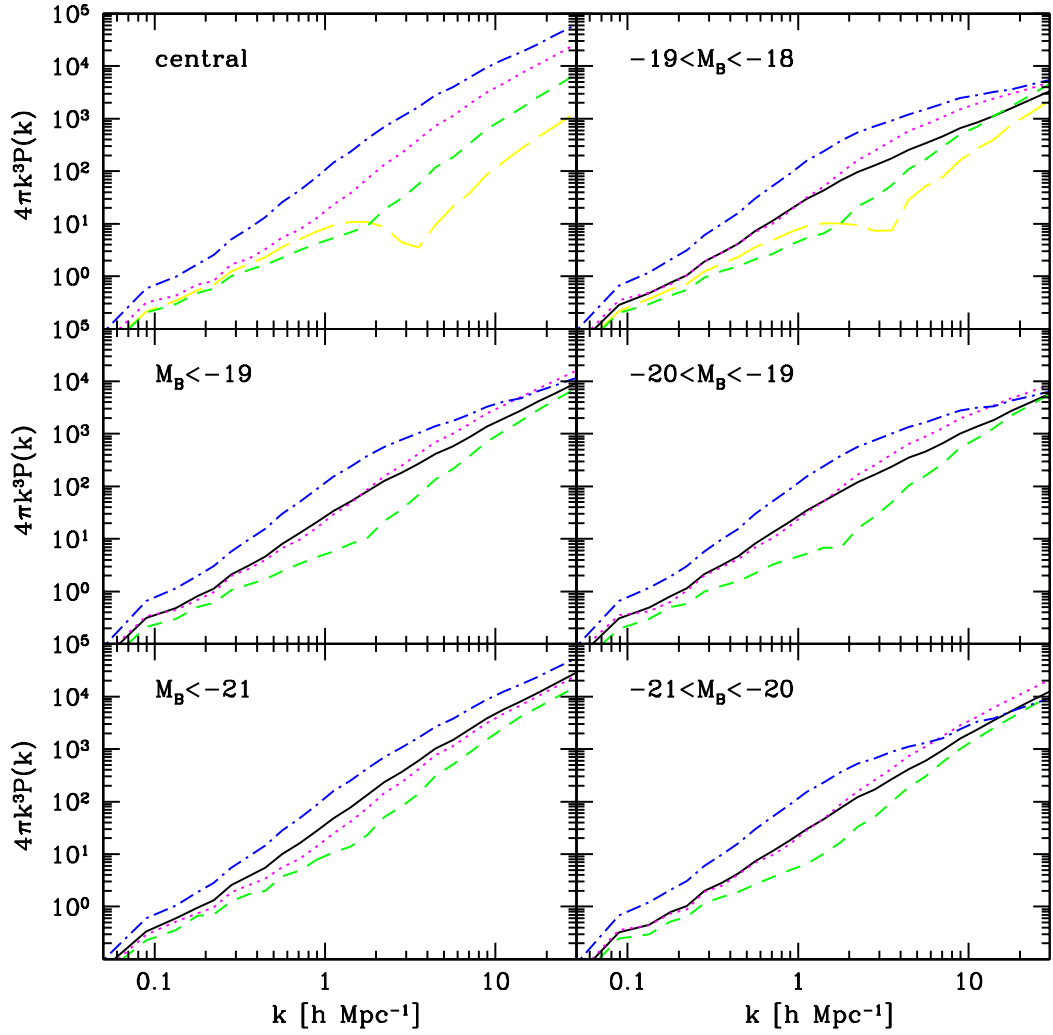, width=150mm}
\caption{
Halo-dark matter cross-power spectra (upper left) for mass
intervals $10^{14}h^{-1} M_{\sun}<M<10^{15}h^{-1} M_{\sun}$,
$10^{13}h^{-1} M_{\sun}<M<10^{14}h^{-1} M_{\sun}$, $10^{12}h^{-1}
M_{\sun}<M<10^{13}h^{-1} M_{\sun}$ and $10^{11}h^{-1}
M_{\sun}<M<10^{12}h^{-1} M_{\sun}$, from top to bottom.
Other plots show the galaxy-dark matter cross-power spectra
for different
galaxy samples as a function of halo mass (similarly to upper left panel) and for all 
galaxies within that sample (solid). 
In the bottom four plots the interval with 
$10^{11}h^{-1} M_{\sun}<M<10^{12}h^{-1}
M_{\sun}$ interval is not shown because there are few or no galaxies 
within the sample in such low mass objects.
}
\label{fig5}
\end{figure*}

\begin{figure*}
\epsfig{file=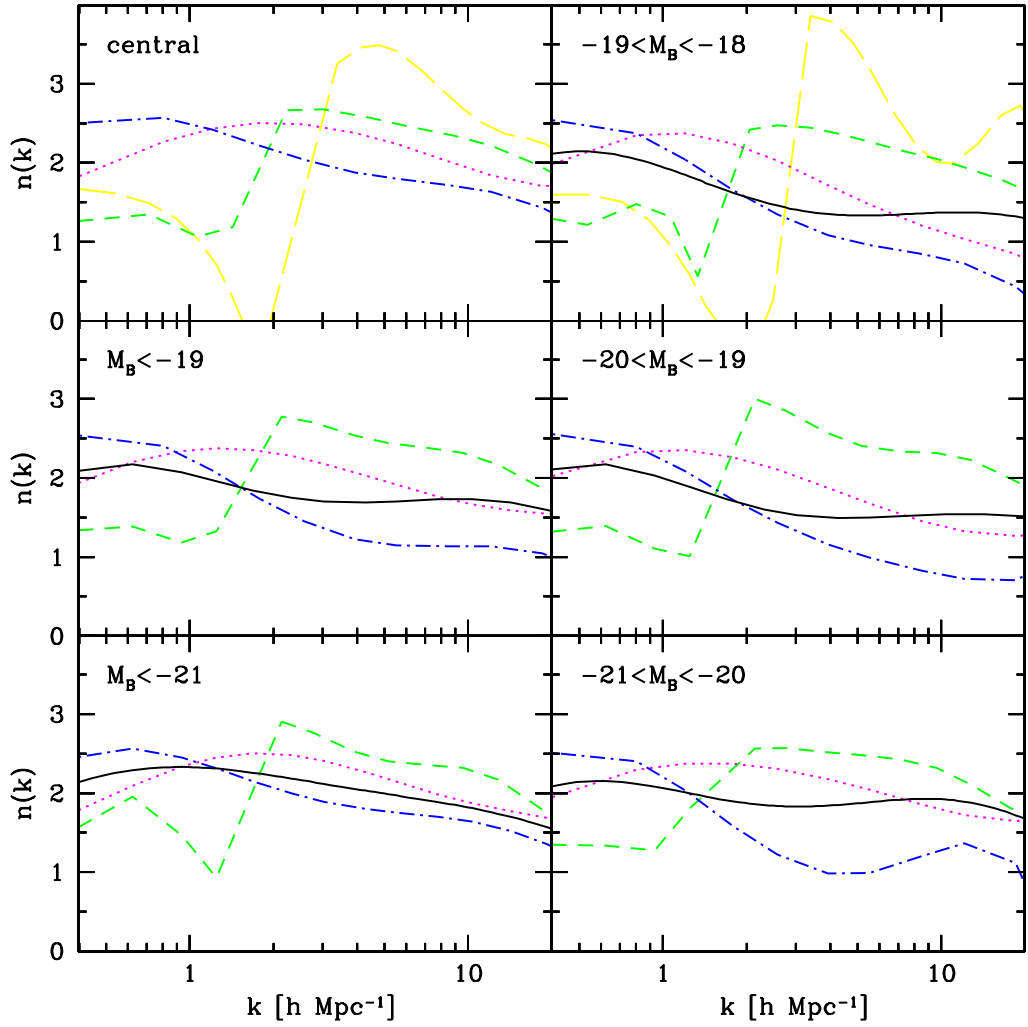, width=150mm}
\caption{
The slope of the cross-power spectra shown in
Fig. \ref{fig5}.  The line types are the same as in Fig. \ref{fig5}. 
}
\label{fig6}
\end{figure*}

\begin{figure*}
\psfig{file=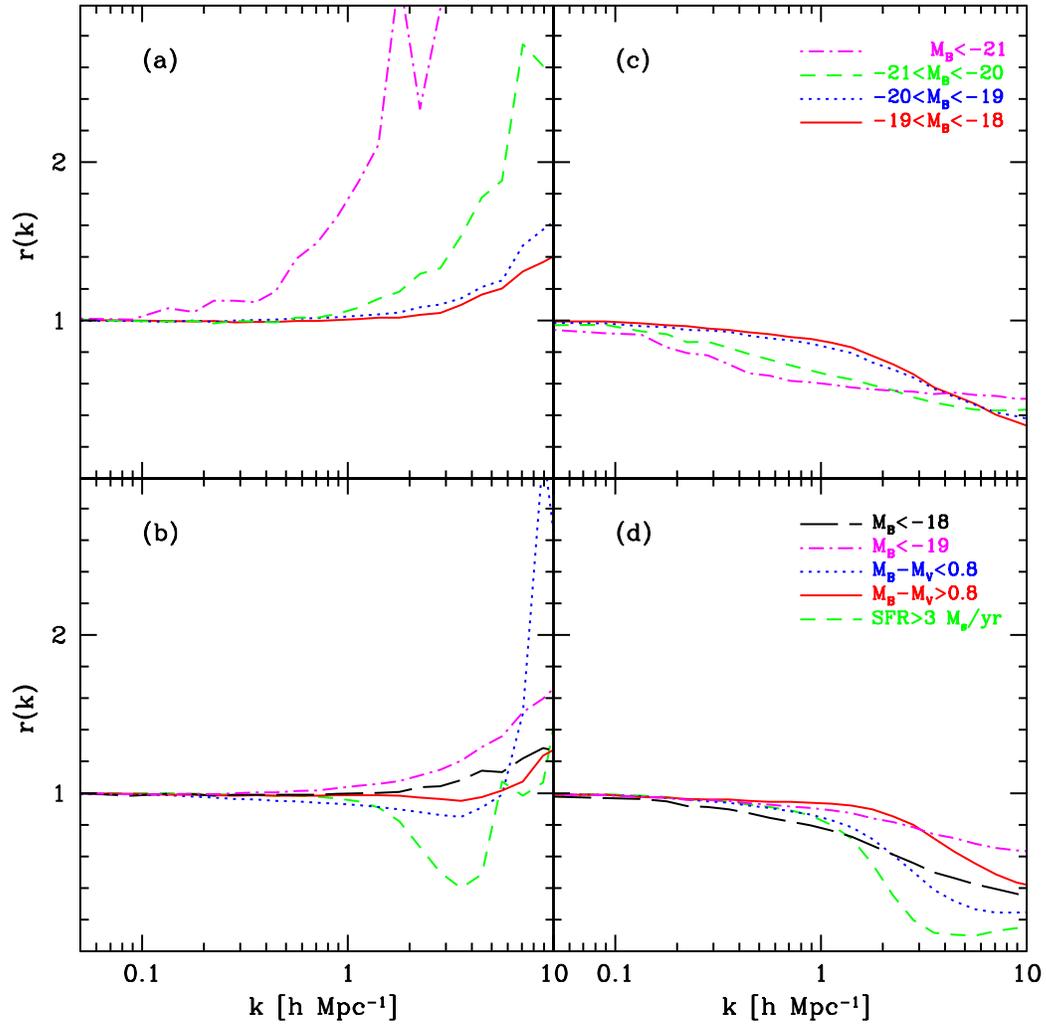, width=150mm}
\caption{
Correlation coefficient $r(k)$ as a function of wavevector $k$
and galaxy sample. In panels (a) and (b) $r(k)$ is derived
from the shot noise
corrected spectra, while in panels (c) and (d) without the
shot noise correction. Galaxy samples are the same as in the Fig. \ref{fig2}.}
\label{fig8}
\end{figure*}

Plots in Fig. \ref{fig5} show the cross-power spectra for several
galaxy types selected by their intrinsic luminosity as a function of 
halo mass. Smaller haloes give lower amplitude of the cross-power spectrum as 
expected. Note that for most galaxy types 
the total cross-spectrum shown with 
a dotted line does not agree with any individual mass range. 
The logarithmic slope $\alpha(k)=d \ln \sigma^2(k) /d\ln k$ 
of the total cross-spectrum is shallower that the 
slope of individual mass intervals (Fig. \ref{fig6}).
On larger scales the cross-spectrum is dominated by the high mass
range of the haloes, while on smaller scales smaller haloes become 
dominant and since these give lower amplitude of the cross-spectrum 
the overall slope is shallower. The only exception are the very bright 
galaxies with $M_B<-21$ which seem to reside predominantly in haloes 
more massive than $10^{13}h^{-1} M_{\sun}<M<10^{14}h^{-1} M_{\sun}$ and their 
cross-correlation function agrees well with the cross-correlation 
for this mass interval.

Second complication which is important for large haloes such as groups 
and clusters is that when there are many galaxies inside the halo 
most of them are not at the halo 
centre. When a galaxy is at the 
outskirts of a halo the lensing effect will on average be reduced. 
If the galaxies are distributed like the dark matter then one can account 
for this effect. 
However, galaxies may not 
be distributed in the same way as the dark matter inside the haloes 
(Diaferio et al. 1999)
and this becomes difficult to model analytically.
This effect is shown by comparing
the first plot of Fig. \ref{fig5}, where the
cross-spectra of central galaxies are shown as a function of halo mass 
intervals, to the other plots in that figure, where all the galaxies
inside the halo are selected. For the two most massive intervals there 
is a significant difference between the central galaxy and all galaxies
cross-spectra. The all galaxy spectra are substantially lower in amplitude as
expected if most galaxies are not central. For lower mass intervals the 
difference between the two is smaller, since such haloes contain
on average at most one bright galaxy, which is usually the 
central one. 

It is instructive to investigate the slope of the cross-spectrum. 
For central galaxies this gives the slope of the dark matter 
profile as a function of scale. On small scales the slope 
approaches $\alpha \sim -1.5$ for the largest haloes with mass range 
$10^{14}h^{-1} M_{\sun}<M<10^{15}h^{-1} M_{\sun}$ for which the force
resolution in units of virial radius is the highest. 
It is possible that the slope would become 
even shallower if higher resolution simulation are used, although 
some of such higher resolution simulations suggest the slope remains close 
to $\alpha \sim -1.5$ down to very small scales (Moore et al. 1999).
This slope steepens
to  $\alpha \sim -2.5$ at the virial radius, beyond which the
contribution from the other haloes becomes important and makes the slope 
less steep again. The behaviour for smaller halo mass intervals 
is similar, but shifted to smaller scales. At the resolution limit 
($k \sim 20 \kunit$) the slope decreases with the decreasing halo mass,
caused by the finite force resolution. For the lowest mass interval the 
transition between the halo profile and other halo correlations is 
quite prominent, but this is at least in part caused by the finite mass
resolution because there are no very small haloes very close
to the virial radius of such haloes.
For all the galaxies, except the very brightest ones, 
the cross-spectrum slope is always shallower
than for the central galaxies when viewed as a function of the halo mass.

\section{Conclusions}
\label{seven}

We have investigated the galaxy-dark matter cross-correlation function 
and power spectrum using the semi-analytic galaxy formation models 
coupled with a high resolution N-body simulation. These cross-correlation 
spectra are the necessary ingredients to interpret the
observed  correlations based on weak lensing and galaxy 
positions. Examples are galaxy-galaxy lensing, QSO-galaxy correlations
and foreground-background galaxy correlations. 
The advantage of the approach used here is that it includes many complicating 
effects such as the variety of the dark matter environments around the galaxies,
the distribution of galaxies inside the dark matter haloes, correlations 
between the galaxies and the dark matter in the neighbouring haloes. The 
limitation of this approach is the mass and force resolution, which 
limits our study to haloes more massive than $10^{11}h^{-1}M_{\sun}$
(limiting the completeness of galaxy sample to $M_B<-18$)
and scales larger than $50 \; h^{-1}{\rm kpc}$ . This can be overcome to some extent 
by using the analytic model, which can extend these results to 
smaller scales (Seljak 2000). 

The cross-correlation spectra on large scales fall between the 
dark matter and galaxy auto-correlation spectra. As a function 
of redshift the correlation strength of bright galaxies 
first decreases to $z \sim 1$
and then remains approximately flat because such galaxies 
become rare and biasing increases above unity, even if the dark matter
correlation strength continues to decrease. 
In terms of the
cross-correlation coefficient we find it 
is close to unity on large scales. This is good 
news for attempts to determine the scale dependent bias and the dark matter 
power spectrum directly from the galaxy correlations 
and the galaxy-galaxy lensing. It breaks down on small 
scales where effects such as the average number of galaxies inside the halo 
dropping below unity and presence of central galaxies cause the 
cross-correlation spectrum to exceed the auto-correlation spectra. 
This happens on larger scales for rarer galaxies, which are therefore 
less suitable for the determination of the dark matter power spectrum. 

While on large scales the cross-correlations reflect clustering of
large scale structure, on small scales they reflect more the dark matter 
environments around galaxies. These have been often interpreted as 
the dark matter profiles of haloes, but such interpretation is complicated
by the effects of multiple galaxies inside clusters and by having 
different halo mases dominate on different scales. For example, the 
observed signal at $1 \runit$ by SDSS (Fischer et al. 2000) 
does not mean that the haloes of galaxies such 
as our own can be observed at such large distances, but rather that 
groups and clusters extend to these distances or that haloes are correlated
with neighbouring haloes.
Still, the predicted signal on small scales 
does increase with the galaxy luminosity, as predicted by 
the Tully-Fisher relation if the dark matter makes a significant 
contribution to the maximal rotation velocity. This shows that 
the cross-correlations can become a complementary tool to study the dark matter 
environments of the galaxies. This can be potentially a powerful 
probe of the dark matter around the galaxies, because it is sensitive to 
larger distances from the centre which are less affected by the 
baryonic component. The upcoming surveys such as SDSS and 2dF should 
be able to extract this information with a high statistical precision, 
making galaxy-galaxy lensing an important
observational tool in connecting galaxies 
to their dark matter environment.

\section*{Acknowledgments}
We thank Guinevere Kauffmann and Antonaldo Diaferio for providing results of GIF N-body
simulations and semi-analytic simulations and for help with them. We also thank 
Phil Fischer for providing us with the galaxy redshift distributions
as measured by the SDSS team.
J.G. was supported by grants 2P03D00618 and 2P03D01417 from 
Polish State Committee for Scientific Research.
U.S. acknowledges the support of NASA grant NAG5-8084.

\newpage

\end{document}